\documentclass[12pt]{elsart}
\usepackage{psfig,epsfig}
\usepackage[usenames]{color}
\journal{Lepton Photon 2007}  

\def\np#1#2#3{    {\it Nucl. Phys. }{\bf #1} (#2) #3}

\def\pl#1#2#3{    {\it Phys. Lett. }{\bf #1} (#2) #3}

\def\pr#1#2#3{    {\it Phys. Rev. }{\bf #1} (#2) #3}

\def\zp#1#2#3{    {\it Zeit. f\"ur Physik }{\bf #1} (#2) #3}

\def\re#1{{\mathrm Re}\left\{#1\right\} }
\def\im#1{{\mathrm Im}\left\{#1\right\} }
\newcommand{\com}[1]{ \par }

\def\evg{\, g_{eV}^\gamma}
\def\tvg{\, g_{\tau V}^\gamma}
\def\evz{\, g_{eV}^Z}
\def\eaz{\, g_{eA}^Z}
\def\tvz{\, g_{\tau V}^Z}
\def\taz{\, g_{\tau A}^Z}
\def\tz{\, g_{\tau T}^Z}
\def\tg{\, g_{\tau T}^\gamma}
\def\evg2{(g_{eV}^\gamma)^2}
\def\tvg2{(g_{\tau V}^\gamma)^2}
\def\evz2{(g_{eV}^Z)^2}
\def\eaz2{(g_{eA}^Z)^2}
\def\tvz2{(g_{\tau V}^Z)^2}
\def\taz2{(g_{\tau A}^Z)^2}
\def\tz2{(g_{\tau T}^Z)^2}
\def\tg2{(g_{\tau T}^\gamma)^2}

\def\re#1{{\mathrm Re}\left\{#1\right\} }
\def\im#1{{\mathrm Im}\left\{#1\right\} }

\def\dps{\displaystyle}
\newcommand{\beq}{\begin{equation}}
\newcommand{\eeq}{\end{equation}}
\newcommand{\bi}{\begin{itemize}}
\newcommand{\ei}{\end{itemize}}
\newcommand{\bea}{\begin{eqnarray}}
\newcommand{\eea}{\end{eqnarray}}
\newcommand{\bes}{\begin{eqnarray*}}
\newcommand{\ees}{\end{eqnarray*}}

\begin{document}
\begin{frontmatter}
\title{$\tau$ electric dipole moment
with polarized beams}
\author[Montevideo]{G. A. Gonz\'alez-Sprinberg},
\author[Valencia]{J. Bernab\'eu} and
\author[Valencia]{J. Vidal}
\address[Montevideo]{Instituto de F\'{\i}sica,
 Facultad de Ciencias, Universidad de la Rep\'ublica,
 Igu\'a 4225, 11400 Montevideo, Uruguay}
\address[Valencia]{Departament de F\'{\i}sica Te\`orica
Universitat de Val\`encia, E-46100 Burjassot,Val\`encia, Spain\\
and\\
IFIC, Centre Mixt Universitat de Val\`encia-CSIC, Val\`encia, Spain}

 \begin{abstract}
High luminosity Super B/Flavor factories, near and
on top of the $\Upsilon$ resonances,
 allow for a detailed investigation of CP-violation in $\tau$ physics. In particular, bounds on the $\tau$
electric dipole moment can be obtained from CP-odd
observables. We perform an
 independent analysis from other low and high energy data.
For polarized electron beam a  CP-odd asymmetry,
associated to the normal polarization term, can be used to set stringent bounds on the
$\tau$ electric dipole moment.

%
\end{abstract}
\end{frontmatter}

\section{Introduction}
The standard model (SM) describes with high accuracy most of the
particle physics experiments \cite{pdg}. However,
 the first clue to physics beyond the SM has been found in neutrino physics \cite{nu}.
This opens  the possibility for  new phenomena
related to  CP violation physics, particularly  in the leptonic sector.
The time reversal odd electric dipole
moment (EDM) for leptons, specially the electron and the muon, has been extensively investigated and strong limits were measured \cite{pdg}:
\beq
d^e_\gamma \,= \,(0.07\pm0.07) \times 10^{-26} {\it e cm}
\eeq
Present  bounds for the $\tau$ lepton EDM are much lower than for the electron or muon case \cite{belle}:
\beq
-0.22 \, \it{ e cm } \,<\, Re (d^\tau_\gamma) \times 10^{16}\, <\,
0.45\, \it{e cm} \;\; ( 95\% \, C.L. ).
\label{taub}
\eeq
The EDM effective
 operator flips
 chirality and,  therefore, the $\tau$ lepton
physics is expected   to be more sensitive to contributions
coming from new
physics.
In the SM the  CP-violation is introduced by the CKM mechanism. There
the EDM and weak-EDM ({\it i.e.} the T-odd diagonal coupling with the Z) are
generated at very high order in the coupling constant. This
opens a way to  test many models: CP-odd observables related to
EDM would give no appreciable effect from the SM and
any experimental signal should be identified with  beyond the
SM physics where the EDM can be generated  at 1-loop. In refs.
\cite{nt,heidel} the $\tau$ weak-EDM has been studied in CP-odd
observables  at high energies through linear
polarizations and spin-spin correlations \cite{l3,ao}. In
ref.\cite{rindani} the sensitivity to the weak-EDM in spin-spin
correlation observables was studied for tau-charm-factories with
polarized electrons. EDM limits for the $\tau$,
from CP-even observables such as total cross sections or decay
widths, have also been considered in \cite{paco,grif,masso}.  The
 limit in Eq.(\ref{taub}) was found by the BELLE Collaboration measuring
  CP-odd spin correlation observables.
Most of the statistics for the $\tau$ pair
production was dominated in the past by LEP,
but the high luminosity of the
B factories and their upgrades  have nowadays the largest $\tau$ pair samples. In the future,
the data will be dominated by the proposed Super B/Flavor
factories \cite{super}. These facilities may also have the
possibility of polarized beams.
In this paper we present new CP-odd observables, related to  the $\tau$ pairs produced at low energies with polarized beams, that may lead to competitive results  with the present
bounds for the $\tau$ EDM.
\section{Effective Lagrangian}
We parametrize the low energy new physics effects by
an effective Lagrangian built with the SM particle spectrum containing higher dimension gauge
invariant operators suppressed by the scale  of new physics,
$\Lambda$ \cite{buch}.  The leading order EDM and weak-EDM Lagrangian for CP violation has only two
dimension six operators \cite{arcadi} that contribute:
\begin{equation}
\label{eq:leff}
\mathcal{ L}_{eff} = i \alpha_B \mathcal{ O}_B + i
\alpha_W \mathcal{ O}_W + \mathrm{h.c.}
\label{eq:interaccio}
\end{equation}
where
$\alpha_B$ and $\alpha_W$ are real couplings and the operators are defined as follows:
\begin{eqnarray}
\label{eq:ob}
\mathcal{O}_B = \frac{g'}{2\Lambda^2} \overline{L_L} \varphi \sigma_{\mu\nu}
 \tau_R B^{\mu\nu} ~,& \hspace*{1cm} \mathcal{ O}_W
 = \frac{g}{2\Lambda^2} \overline{L_L} \vec{\tau}\varphi
\sigma_{\mu\nu} \tau_R \vec{W}^{\mu\nu}  ~.
\end{eqnarray}
Here $L_L=(\nu_L,\tau_L)$ is the $\tau$ leptonic doublet,
$\varphi$  is the Higgs doublet, $B^{\nu\nu}$ and $\vec{W}^{\mu\nu}$ are the
U(1)$_Y$ and SU(2)$_L$ field strength tensors, and $g'$ and $g$ are the gauge
couplings.
After spontaneous symmetry breaking  the interactions in Eq.(\ref{eq:interaccio})
produce the usual EDM effective operators:
\begin{eqnarray}
\mathcal{ L}_{eff}^{\gamma, Z} &=& - i  \,d^\tau_\gamma
\,\overline{\tau} \sigma_{\mu\nu}
 \gamma^5 \tau F^{\mu\nu} -
 i  \,d^\tau_Z \,\overline{\tau} \sigma_{\mu\nu}
\gamma^5 \tau
Z^{\mu\nu}
\label{eq:leff_fin}
\end{eqnarray}
where $F_{\mu\nu}=\partial_\mu A_\nu-\partial_\nu A_\mu$
  and
$Z_{\mu\nu}=\partial_\mu Z_\nu-\partial_\nu Z_\mu$
are  the abelian field strength tensors of the photon and
  $Z$ gauge boson and $d^\tau_\gamma$ and $\d^\tau_Z$ are the electric and
  weak-electric dipole moments, respectively.
Note that as long as $
|q^2| \ll \Lambda^2$ there is no need to distinguish between the
 new physics contribution  to the  EDM  form factors and the EDM dipole moment.

\begin{figure}[hbtp]
\begin{center}
\epsfig{file=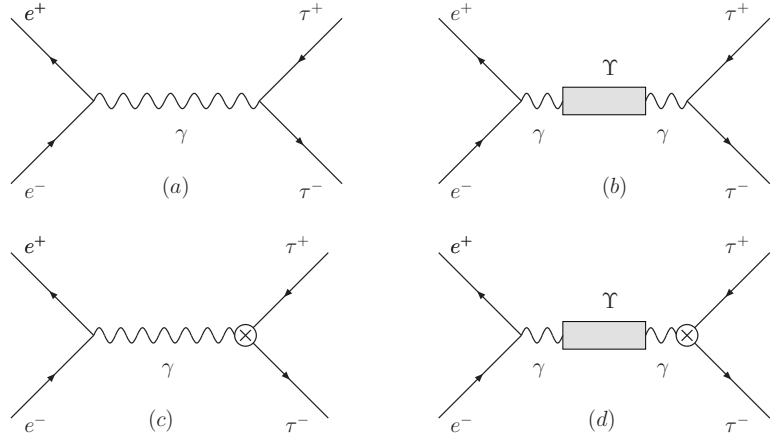,width=0.75\textwidth}
\end{center}
\caption{Diagrams (a) direct $\gamma$ exchange (b) $\Upsilon$
production (c) EDM in $\gamma$ exchange (d) EDM at the
$\Upsilon$-peak.} \label{fig:figura1}
\end{figure}

The $e^+\, e^- \longrightarrow
\tau^+ \tau^-$ cross section has contributions coming from
the SM and the terms in
Eq.(\ref{eq:leff_fin}).
Tree level contributions come from
  $\gamma$
or $\Upsilon$
exchange in the s-channel and are shown in Fig.(\ref{fig:figura1}). Other contributions coming from
diagrams where at least one photon line is substituted by a $Z$
are suppressed by powers of $\left(q^2/M_Z^2\right)$.

As stated in the introduction the CKM contributions to CP-odd observables are far below
 the present experimental sensitivity.
The  bounds on the EDM that one may get are just the ones coming
from  new physics.
\section{Low energy polarized beams and the EDM.}
For
longitudinally polarized electrons
the $\tau$-EDM  modifies the angular
distribution for the $e^+e^-  \longrightarrow
\tau^+(s_+)\tau^-(s_-)$  cross section.
The normal -to the scattering plane- polarization ($P_N$) of each $\tau$
 is the only linear
component which is $T$-odd. For CP-conserving interactions, the
CP-even term $(s_++s_-)_N$ of the normal polarization only gets
contribution through the combined
effect of both an helicity-flip transition and the presence of
absorptive parts, which are both
suppressed in the SM. For a CP-violating interaction,
such as an EDM, the $(s_+-s_-)_N$ CP-odd term  gets a
non-vanishing value without the need of absorptive parts.
$P_N$ is also even under parity ($P$) so any observable
sensitive to the EDM will need in addition
a $P$-odd contribution. In our case this comes  from the  longitudinally polarized
electrons.
We use the notation of references \cite{arcadi,nos,nos07}.
The $\mbox{\boldmath $s$}^\pm$  are the $\tau^\pm$  spin vectors in
the $\tau^\pm$ rest system, $s_\pm=(0, s^x_\pm, s^y_\pm, s^z_\pm)$.
Polarization
along the directions
$x,y,z$ are called transverse
 (T), normal (N) and longitudinal (L), respectively.
Let us   first consider the s-channel $\tau$-pair production (diagrams (a) and (c)
in Fig.(\ref{fig:figura1})).
We  assume  that the $\tau$ production plane and
direction of flight can be fully reconstructed, which is the case
 for both $\tau$'s decaying semileptonically \cite{kuhn}.
The spin dependent part of the differential cross section for $\tau$ pair production with
polarized electrons with helicity $\lambda$  is:
%
\bea
\left.\frac{d \sigma^{S}}{d
\Omega_{\tau^-}}\right|_\lambda=\frac{\alpha^2}{16\ s}\, \beta &&
\left\{\lambda\left[
(s_-+s_+)_xX_++(s_-+s_+)_zZ_++(s_--s_+)_yY_- \right)]+\right.\nonumber \\
&&\quad \left. (s_--s_+)_x X_- + (s_--s_+)_z Z_-\right\}
\eea
where
\bea
X_+&=&\frac{1}{\gamma}\sin\theta_{\tau^-},\qquad\qquad\qquad\qquad
X_-=-\frac{1}{2}\sin(2\theta)\,
\dps\frac{2m_\tau}{e}\im{d_\tau^\gamma},\nonumber\\
Z_+&=&+\cos \theta_{\tau^-},\qquad\qquad\qquad\qquad
Z_-=-\frac{1}{\gamma}\sin^2\theta\,
\dps\frac{2m_\tau}{e}\im{d_\tau^\gamma},\nonumber\\
Y_-&=&\gamma\beta\, \sin \theta_{\tau^-}\,
\dps\frac{2m_\tau}{e}\re{d_\tau^\gamma}. \label{cross3}
\eea
and $\alpha$ is the fine structure constant, $s=q^2$ is the squared CM energy
and $\gamma = \sqrt{s} / 2 m_\tau$,
$\beta = \left( 1 - 1 / \gamma^2 \right)^{1/2}$ are the dilation factor and
$\tau$ velocity, respectively.
  Eq.(\ref{cross3}) shows that the $\tau$-EDM is the leading
contribution to the Normal Polarization.

The cross section for the process
$ e^+e^-(pol) \rightarrow \gamma \rightarrow \tau^+\tau^-
\rightarrow h^+\bar{\nu}_\tau h^-\nu_\tau
$
%
can be written as \cite{tsai}:
\begin{eqnarray}
&&\hspace*{-1cm}d\sigma \left.\left(e^+e^-\rightarrow \gamma
\rightarrow \tau^+\tau^- \rightarrow h^+\bar{\nu}_\tau h^-\nu_\tau\right)\right|_\lambda=
4\, d\sigma
\left.\left(e^+e^- \rightarrow \tau^+(\overrightarrow{n}_+^*)
\, \tau^-(\overrightarrow{n}_-^*)\right)\right|_\lambda\nonumber \\
&&\times \, Br(\tau^+ \rightarrow h^+\bar{\nu}_\tau)
Br(\tau^- \rightarrow h^-\nu_\tau)
\frac{d\Omega_{h^+}}{4\pi}\, \frac{d\Omega_{h^-}}{4\pi}
\label{eq:cros1}\end{eqnarray}
with
\beq \overrightarrow{n}_\pm^*= \mp\alpha_\pm
\frac{\overrightarrow{q}^{  *}_\pm}{
\arrowvert\overrightarrow{q}^{  *}_\pm\arrowvert} =
\mp\alpha_\pm(\sin\theta_{\pm}^*\, \cos\phi_\pm,
\sin\theta_{\pm}^*\, \sin\phi_\pm,\cos\theta_{\pm}^*)\nonumber\\
\eeq
and  $\theta_{\tau^-} $ is the center of mass angle
of the $\tau^-$ with respect to the electron,
$\phi_{\pm}$ and  $\theta^*_{\pm}$ are the azimuthal and polar angles of the
produced hadrons $h^\pm$ ($\hat{q}^*_{\pm}$) in the $\tau^\pm$
rest frame (the * means that the quantity is
 given in the $\tau$ rest frame).
If we integrate  over the $\tau^-$ angular variables then all
the information on the $Z_+$ and $X_-$ terms of
the cross section is eliminated:
\begin{eqnarray}
\left.d^4\sigma^{S}\right|_\lambda &=&
\frac{\pi^2\alpha^2\beta}{4\, s}
\, Br(\tau^+ \rightarrow h^+\bar{\nu}_\tau)
Br(\tau^- \rightarrow h^-\nu_\tau) \, \frac{d\Omega_{h^+}}{4\pi}\,
\frac{d\Omega_{h^-}}{4\pi}\, \times \nonumber\\
&&\Bigg\{\frac{\lambda}{\gamma}\, \left[(n_-^*)_x+(n_+^*)_x\right] +\lambda\,
\gamma\beta\, \left[(n_-^*)_y-(n_+^*)_y\right]\frac{2 m_\tau}{e}\,
\re{d^\gamma_\tau}\\
&&+\frac{4}{3\gamma}\left[(n_-^*)_z-(n_+^*)_z\right]\frac{2 m_\tau}{e}\,
\im{d^\gamma_\tau}\Bigg\} \label{eqxy1}
\end{eqnarray}
Subtracting  for
different helicities leaves only the real part of the $\tau$-EDM:
\beq
\left.\d^2\sigma^{S}\right|_{Pol( e^-)}\equiv \left.d^4\sigma^{S}\right|_{\lambda=1}-
\left.d^4\sigma^{S}\right|_{\lambda=-1}\label{spol1}
\eeq
Keeping only azimuthal angles and integrating all
other variables one gets:
\bea
\left.\frac{d^2\sigma^{S}}{d\phi_- d\phi_+}\right|_{Pol( e^-)}
&=&   \frac{\pi\alpha^2\beta}{32 s}\,
Br(\tau^+ \rightarrow h^+\bar{\nu}_\tau)
Br(\tau^- \rightarrow h^-\nu_\tau)
  \times  \nonumber\\
 & & \Big\{\frac{1}{\gamma}\left[(\alpha_-) \cos\phi_--(\alpha_+)\cos\phi_+\right]+
 \label{eqxy3} \\
&&\beta\, \gamma  \left[(\alpha_-) \cos\phi_--(\alpha_+)\cos\phi_+\right]\frac{2
m_\tau}{e}\,\re{d^\gamma_\tau}  \Big\}
\eea
We can now define the
azimuthal asymmetry as:
\begin{equation}
A_N^{\mp} =\frac{\sigma^\mp_{L} - \sigma^\mp_{R}}{\sigma}=
\alpha_\mp\frac{3\pi\gamma\beta}{8(3-\beta^2)}\frac{2
m_\tau}{e}\,\re{d^\gamma_\tau}\label{asym}
\end{equation}
where
\begin{eqnarray}
\sigma^\mp_L &=& \int_0^{2\pi}d\phi_\pm \left[\int_0^\pi d\phi_\mp\,\left.
\frac{d^2\sigma^S}{d\phi_- d\phi_+}\right|_{Pol (e^-)}\right]=\nonumber\\
&&\quad Br(\tau^+ \rightarrow h^+\bar{\nu}_\tau)
Br(\tau^- \rightarrow h^-\nu_\tau)\,\alpha_\mp\frac{(\pi\alpha\beta)^2\gamma}{8s}\,
\frac{2m_\tau}{e}\,\re{d^\gamma_\tau}\\
\sigma^\mp_R &=& \int_0^{2\pi}d\phi_\pm \left[\int_\pi^{2\pi}d\phi_\mp\,\left.
\frac{d^2\sigma^S}{d\phi_- d\phi_+}\right|_{Pol (e^-)}\right]=\nonumber \\
&&\quad -Br(\tau^+ \rightarrow h^+\bar{\nu}_\tau)
Br(\tau^- \rightarrow h^-\nu_\tau)\,\alpha_\mp\frac{(\pi\alpha\beta)^2\gamma}{8s}\,
\frac{2m_\tau}{e}\,\re{d^\gamma_\tau}
\end{eqnarray}
All other terms in the  cross
section are eliminated when we integrate in this way.
Notice that this integration procedure does not erase
suppressed contributions coming from the CP-even term of the Normal
Polarization.
To eliminate spurious CP-even terms we define a CP-odd observable
by summing up the
asymmetry in Eq.(\ref{asym}) for $\tau^+$ and for $\tau^-$
\beq
A_N^{CP}=\frac{1}{2}\left(A_N^++A_N^-\right)= \alpha_h
\frac{3\pi\gamma\beta}{8(3-\beta^2)}\frac{2
m_\tau}{e}\,\re{d^\gamma_\tau}\label{asimcp}
\eeq
The $\gamma - Z$ interference has been considered in ref.\cite{nos07}  at $q^2 = (10 \,GeV)^2$;
 this contribution is suppressed by a factor of the order of  $10^{-6}$. This is  two
 orders of magnitude below the expected sensitivity for the asymmetries. In any case these
terms do not contribute to the CP-odd $A_N^{CP}$ of Eq.(\ref{asimcp}).

All these ideas can be applied for $e^+e^-$ collisions at the $\Upsilon$ peak
where the $\tau$ pair production is  mediated by the $\Upsilon$ resonances:
$e^+e^- \rightarrow \Upsilon \rightarrow \tau^-\tau^-$.
In this case the resonant diagrams (b) and (d) of Fig.(1) dominate the
process on the $\Upsilon $ peaks. This has been extensively
discussed in ref.\cite{Bernabeu:2004ww}.
The main result is that the $\tau$ pair production at the $\Upsilon$
peak introduces the same $\tau$ polarization matrix terms as the
direct production with $\gamma$ exchange (diagrams (a) and (c)) except for a
 the overall multiplicative factor $|H(s)|^2$  in the cross
section:
\beq H(M_\Upsilon^2) = -i \, \frac{3}{\alpha}
Br\left(\Upsilon \rightarrow e^+e^-\right) \label{factor}
\eeq
Besides, at the $Upsilon$ peak,  the interference of diagrams
(a) and (d) plus the interference of diagrams (b) and (c) is
exactly zero and so it is the interference of diagrams (a) and
(b). Thus, the only contributions proportional to the   EDM
come  with the interference of diagrams (b) and (d), while diagram
(b) squared gives the leading contribution to the cross section.
Finally we obtain no changes in the asymmetries we have already computed:
 the only difference  is in the value of the resonant production cross section
at the $\Upsilon$ peak that is multiplied by the overall factor
$|H(M_\Upsilon^2)|^2$.
\section{Bounds on the EDM }
Let us discuss the $\tau$-EDM bounds  that can be set by measuring this observable.
We assume a set of
integrated luminosities  for high statistics $B$/Flavor factories.
We also consider the decay channels $\pi^\pm \; \bar{\nu_\tau}$ or $\rho^\pm\;
\bar{\nu_\tau}$ ({\it i.e.} $h_1 , \;h_2= \pi ,\; \rho$) for the traced $\tau^\pm$,
while we sum up over $\pi^\mp \; \nu_\tau$ and $\rho^\mp\;
  \nu_\tau$ hadronic decay channels
for the non traced $\tau^\mp$.

The  bounds for the
$\tau$-EDM that can be set in different scenarios are:
\begin{eqnarray}
\hspace*{-.8cm}|\re{d^\gamma_\tau}|&\le& 4.4\ 10^{-19}\ e cm,  \, \mbox{Babar + Belle at $2 ab^{-1}$}\nonumber\\
\hspace*{-.8cm}|\re{d^\gamma_\tau}|&\le& 1.6\ 10^{-19}\ e cm, \,
\mbox{SuperB/Flavor  factory, 1 yr running, $15 ab^{-1}$}\nonumber\\
\hspace*{-.8cm}|\re{d^\gamma_\tau}|&\le& 7.2\ 10^{-20}\ e cm,  \,
\mbox{SuperB/Flavor factory, 5 yrs running, $75 ab^{-1}$}
\end{eqnarray}
These limits improve the PDG ones of Eq.(\ref{taub}) by two orders of
magnitude.

To conclude, we have shown that low energy data allows a clear
separation of the
effects coming from the electromagnetic EDM, the weak EDM and
interference effects. Polarized electron beams open the possibility to put
bounds on the $\tau$ EDM coming from single $\tau$ polarization
observables. These observables allow for an
independent analysis of the EDM bounds from what has been done
with other high and low energy data. The new bounds may by  two orders of magnitude below the
 PDG limits.
\begin{ack}
This work has been supported by CONICYT-PDT-54/94-Uruguay, by MEC
and FEDER, under the grants FPA2005-00711 and FPA2005-01678, and
by Gene\-ralitat Valenciana under the grant GV05/264.
\end{ack}

\end{document}